\documentstyle[12pt,aaspp4]{article}
\begin{document}
\newcommand{\up}[1]{\ifmmode^{\rm #1}\else$^{\rm #1}$\fi}
\newcommand{\zdot}{\makebox[0pt][l]{.}}
\newcommand{\upd}{\up{d}}
\newcommand{\uph}{\up{h}}
\newcommand{\upm}{\up{m}}
\newcommand{\ups}{\up{s}}
\newcommand{\arcd}{\ifmmode^{\circ}\else$^{\circ}$\fi}
\newcommand{\arcm}{\ifmmode{'}\else$'$\fi}
\newcommand{\arcs}{\ifmmode{''}\else$''$\fi}

\title{The Araucaria Project. The Distance to the Sculptor 
dwarf spheroidal galaxy from infrared photometry of RR Lyrae stars
\footnote{Based on observations obtained with the ESO NTT for 
programme 074.D-0318(B)}
}
\author{Grzegorz Pietrzy{\'n}ski}
\affil{Universidad de Concepci{\'o}n, Departamento de Fisica, Astronomy
Group,
Casilla 160-C,
Concepci{\'o}n, Chile}
\affil{Warsaw University Observatory, Al. Ujazdowskie 4, 00-478, Warsaw,
Poland}
\author{Wolfgang Gieren}
\affil{Universidad de Concepci{\'o}n, Departamento de Fisica, Astronomy Group,
Casilla 160-C, Concepci{\'o}n, Chile}
\authoremail{wgieren@astro-udec.cl}
\author{Olaf Szewczyk}
\affil{Universidad de Concepci{\'o}n, Departamento de Fisica, Astronomy
Group,
Casilla 160-C,
Concepci{\'o}n, Chile}
\affil{Warsaw University Observatory, Al. Ujazdowskie 4, 00-478, Warsaw,
Poland}
\author{Alistair Walker}
\affil{Cerro Tololo Inter-American Observatory, Casilla 603, La Serena,
Chile}
\authoremail{awalker@ctio.noao.edu}
\author{Luca Rizzi}
\affil{Joint Astronomy Centre, 66 N. A'ohoku Pl., Hilo, Hawaii, 96720}
\authoremail{luca@ifa.hawaii.edu}
\author{Fabio Bresolin}
\affil{Institute for Astronomy, University of Hawaii at Manoa, 2680 Woodlawn 
Drive, 
Honolulu HI 96822, USA}
\authoremail{bresolin@ifa.hawaii.edu}
\author{Rolf-Peter Kudritzki}
\affil{Institute for Astronomy, University of Hawaii at Manoa, 2680 Woodlawn 
Drive,
Honolulu HI 96822, USA}
\authoremail{kud@ifa.hawaii.edu}
\author{Krzysztof  Nalewajko}
\affil{Warsaw University Observatory, Al. Ujazdowskie 4, 00-478, Warsaw,
Poland}
\authoremail{nalewajk@astrouw.edu.pl}
\author{Jesper Storm}
\affil{Astrophysikalisches Institut Potsdam, An der Sternwarte 16, D-14482
Potsdam, Germany}
\authoremail{jstorm@aip.de}
\author{Massimo Dall'Ora}
\affil{INAF, Osservatorio Astronomico di Capodimonte, I-80131 Napoli, Italy}
\authoremail{dallora@oacn.inaf.it}
\author{Valentin Ivanov}
\affil{European Southern Observatory, Ave. Alonso de Cordova, Casilla
19001, Santiago 19, Chie }
\authoremail{vivanov@eso.org}

\begin{abstract}
We have obtained single-phase near-infrared magnitudes in the J and K bands
for a sample of 78 RR Lyrae stars in the Sculptor dSph galaxy. Applying different
theoretical and empirical calibrations of the period-luminosity-metallicity
relation for RR Lyrae stars in the infrared, we find consistent results and
obtain a true, reddening-corrected distance modulus of 19.67 $\pm$ 0.02 (statistical)
$\pm$ 0.12 (systematic) mag for Sculptor from our data. This distance value is
consistent with the value of 19.68 $\pm$ 0.08 mag which we obtain from earlier V-band
data of RR Lyrae stars in Sculptor, and the V magnitude-metallicity calibration
of Sandage (1993). It is also in a very good agreement with the results obtain 
by Rizzi (2002)  based on tip of the red giant branch  (TRGB, 19.64 $\pm$ 0.08 mag) and
horizontal branch (HB, 19.66 $\pm$ 0.15 mag).

\end{abstract}


\section{Introduction}
The main goal of the Araucaria project is to improve the calibration 
of the cosmic distance scale from accurate observations of the various 
primary stellar distance indicators in nearby galaxies (e.g. Gieren et al. 2005b).
In particular, we are significantly improving the distance
measurements to the Araucaria target galaxies using IR photometry, which minimizes both 
the influence of interstellar extinction, and of metallicity and/or age 
on the brightness of stellar distance indicators. Indeed, in  our 
previous papers we have demonstrated that the red clump stars  
(Pietrzynski and Gieren 2002; 
Pietrzynski, Gieren and Udalski 2003), and Cepheid variable stars 
(e.g. Pietrzynski et al. 2006; Gieren et al. 2005a, 2006, 2008; Soszynski et al. 2006)
are outstanding distance indicators in the near-infrared domain. 

Thanks to several recent and important theoretical and empirical studies of RR Lyrae stars 
we have now good reasons to believe that RR Lyrae stars are also capable of providing 
much improved distances from their magnitudes in the near-infrared regime,
as compared to the traditional optical studies.
Longmore et al. (1986) were the first to show that RR Lyrae stars 
follow a period-luminosity (PL) relation in the K band. 
Then important contributions were added by Carney et al. (1992),  Nemec et al. (1994), Jones et
al. (1993) Frolov and Samus (1998) and Bono et al. (2001).  
Dall'Ora et al. (2004) later demonstrated
that such a relation could be very tight for RR Lyrae stars in a globular cluster,
with its very low metallicity spread. 
Detailed theoretical studies of the RR Lyrae period-mean magnitude-metallicity relations
 in different near-infrared
passbands were done by Bono et al. (2003), and Catelan et al. (2004). 
Sollima et al. (2006) analyzed near-infrared K-band data 
of RR Lyrae stars for some 15 Galactic globular clusters and provided 
the first extensive empirical calibration of the
period-luminosity-metallicity (PLZ) relation in the K band.
Most recently, the zero point of this relation was improved 
based on accurate near-infrared observations of the RR Lyrae star
(Sollima et al. 2008). 
These existing theoretical and empirical results have demonstrated that
a near-infrared PLZ relation for RR Lyrae stars appears to be a very accurate
means for distance determination to galaxies containing an abundant old
stellar population. We are therefore including this tool in our project
and expect that we will be able to improve on its current calibration by
conducting near-infrared observations of large numbers of RR Lyrae variables
in a number of nearby galaxies, and comparing the resulting RR Lyrae distances
to those emerging from other methods, like the TRGB technique.
 
With its rich population of both RRab and RRc stars
which span a wide range of metallicities (Clementini et al. 2005), the Sculptor 
dwarf spheroidal galaxy is an
ideal laboratory for studying the infrared PLZ relations defined by these variables. Therefore we
included this galaxy in the target list of  
the Araucaria project. In this paper, we present first results from 
deep single-phase near-infrared imaging of 78 RR Lyrae stars in Sculptor. We will
show that these data allow a distance determination for Sculptor with very small
intrinsic error. 

\section{Observations, Data Reduction and Calibration}
The near infrared data presented in this paper were collected as a
part of the Araucaria Project, with the ESO NTT 
telescope on La Silla equipped with the SOFI infrered camera
(Moorwood, A., Cuby, J.G., Lindman, C., 1998)
. The 
Large Field setup, with a field of view of 4.9 x 4.9 arcmin, and a
scale of 0.288 arcsec/pixel was used. During one photometric night 
we observed two fields in Sculptor, 
which contain a large number of RR Lyrae stars.
Their locations are shown in Fig. 1. We secured single deep J and Ks 
observations for each field under excellent (0.6") seeing conditions.
In order to account for rapid sky level variations in the infrared
domain, our observations were performed with a dithering technique.
The resulting total integration times were 44 and 21 minutes for 
Ks and J band images, respectively. 

All reductions and calibrations were performed  with a pipeline 
developed and used in the course of the Araucaria Project. 
Briefly, the sky subtraction was applied in two-step process 
implying  masking of stars with the XDIMSUM IRAF package
(see Pietrzy{\'n}ski and Gieren (2002) for more details). 
Then the single images were flat-fielded and stacked into the
final, deep images. PSF photometry and aperture corrections 
were derived using DAOPHOT and ALLSTAR programs, in an identical
manner as described in Pietrzy{\'n}ski, Gieren, and Udalski (2002).  

In order to calibrate our photometry to the standard system 
9 standard stars from the UKIRT list (Hawarden et al. 2001)
were observed under photometric conditions at different 
airmasses together with our target fields. 
The chosen standards have colors which bracket 
the colors of the RR Lyrae stars in Sculptor. Given the large 
number of standard stars the accuracy of the zero point of our 
photometry was estimated to be about 0.02 mag. 

In order to make an external check of our photometry the magnitudes of
the bright stars in our fields were transformed onto the 2MASS photometric system 
and then  compared with the 2MASS photometry. We found 
the following differences (in the sense 2MASS photometry minus our
results): -0.01 $\pm$ 0.11 mag (K), -0.03 $\pm$ 0.13 mag (J).

The J and K band magnitudes of the observed RR Lyrae stars, together with 
the corresponding errors as returned by DAOPHOT, and the time of the beginning
of each exposure are given in Table 1. 

\section{Near-Infrared Period-Luminosity Relations}
Since the IR light curves of RR Lyrae stars are nearly sinusoidal and 
have relatively small amplitudes (e.g. Del Principe et al. 2006), single-phase
measurements of J and K magnitudes of RR Lyrae stars give approximations of reasonable accuracy
to their mean magnitudes in these bands. The PL relations 
in J and K we obtain from our random single-phase data are presented in Fig. 2.
Two sequences are clearly visible, corresponding to the first overtone (RRc) and fundamental 
mode (RRab) pulsators, respectively.
 The relatively large scatter observed in this Figure is caused by the three following factors: 
Firstly, a single phase IR measurement is expected to approximate the mean magnitude of 
a RR Lyrae variable
to about 0.15 mag only. Secondly, a rather large spread of metallicities  
is observed  for the RR Lyrae stars of this galaxy  (Clementini et al. 2005, Kaluzny et al.
1995). Clementini et al (2005) determined a mean metallicity (on the 
Zinn and West scale) of 
RR Lyrae stars of -1.83 $\pm$ 0.26 dex. Assuming a metallicity dependence 
in the K and J bands  of 0.175 and 0.190 mag/dex,
respectively (Catelan et al. 2004),  we expect 
an additional scatter in the IR magnitudes of Fig. 2 of about 0.05 mag (one $\sigma$) 
due to the metallicity inhomogeneity of our sample of RR Lyrae stars in Sculptor.
Finally, the accuracy of our individual magnitudes is between 0.04 and 0.07 mag
for stars having K band magnitudes in the range from 18.6 to 19.5. Clearly, the
largest source for the scatter in Fig. 2 is the replacement of mean magnitudes
derived from complete light curves by the single-phase observations.
Unfortunatelly due to a very large gap between the optical survey by 
Kaluzny et al. (1995) and our IR observations it is not possible to use
the template light curves provided by Jones et al. 
(1996) to calculate more accurate mean magnitudes.

The slopes of the PL relations followed by the RRab and RRc stars, and by
the whole, combined RRab+RRc sample of RR Lyrae stars as obtained from free least-square fits 
to a line 
in both J and K band filters are given in Table 2. In order to merge the first overtone and
fundamental pulsators, a value of 0.127
was added to the logarithm of the periods of the RRc stars, following Dall'Ora et al. (2004). 
Unfortunately, the relatively large errors associated with these slope
determinations do not allow us to conduct a detailed discussion on 
these results. We can only conclude that within the limited accuracy 
we can achieve in this study we cannot detect any significant difference 
between the slopes of the 
K- and J-band PL relations obtained for the RRc and RRab stars,
respectively. The  slopes agree furthermore very well with both, the existing
theoretical (Bono et al. 2003, Catelan et al. 2004)  and empirical (Sollima et
al. 2008) slope results obtained for field RR Lyrae stars.

\section{The Distance Determination}
In order to derive the apparent  distance moduli to the Sculptor galaxy
from our data, we used the following calibrations of the 
near-infrared PL relations of mixed population RR Lyrae stars: \\

 $ {\rm M}_{\rm K}$ =  -1.07 $-$ 2.38 log P + 0.08 [Fe/H] ;      Sollima et al (2008)
 (1) \\

$ {\rm M}_{\rm K}$ = -0.77  $-$ 2.101  logP + 0.231  [Fe/H] ;   Bono et al. (2003)   (2) \\

$ {\rm M}_{\rm K}$ = -0.597 $-$ 2.353 logP + 0.175  logZ   ;   Catelan  et al. (2004)
  (3)\\

$ {\rm M}_{\rm J}$ = -0.141 $-$ 1.773  logP + 0.190  logZ  ;    Catelan  et al. (2004)
 (4)\\

We recall that the calibration of Sollima et al. (2008)
was constructed for the 2MASS photometric system, while 
the calibrations of Bono et al. and Catelan et al. are valid 
for the Bessel and Brett and Glass systems. Therefore,
we transformed our own data, calibrated onto the UKIRT system (Hawarden et
al. 2001), to the Bessel and Brett and Glass  systems using the transformations given by
Carpenter (2001), before calculating distances using the calibrations 
of Bono et al. and Catelan et al., respectively.
Since there is virtually no difference between 
the K-band filters of 2MASS and UKIRT (Carpenter 2001), 
we did not apply any transformations to our data 
for the Sollima et al. calibrations.

Assuming the mean metallicity of our RR Lyrae sample to be
 -1.83 $\pm$ 0.26 dex (Clementini et al.
2005), we calculated the K- and J- band apparent distance moduli as
tabulated in Table 3. The corresponding fits to a straight line
are presented in Figure 3. The calculations were made for the RRab star sample alone,
 and for the combined sample of RRab and RRc stars, after adding 0.127 in
logarithm of period to the RRc stars. The results 
obtained for both samples are in an excellent agreement 
(within 0.5 $\sigma$), so hereinafter we decided to adopt the slightly
more accurate results from the whole sample of RR Lyrae stars observed in this
galaxy (e.g. RRab and RRc) for further discussion.

To correct the obtained apparent distance moduli for interstellar reddening,
we adopted a foreground reddening of E(B-V) = 0.018 mag which is  calculated toward 
the Sculptor galaxy from the  Galactic exctinction
maps of Schlegel et al. (1998). Assuming the reddening law from this same
paper we obtain the following  values for the selective extinctions in the
different bands: $A_{K}$=0.007 mag, $A_{J}$=0.017 mag, $A_{V}$=0.059
mag. The resulting true distance moduli for Sculptor are summarized in Table 4. 
We also include the true distance modulus to the Sculptor 
galaxy from the optical V band data of Kaluzny et al. (1995) in this Table.

\section{Discussion}
The results in Table 4 demonstrate that the true distance moduli obtained from the 
theoretical calibrations of the RR Lyrae PLZ relation in the J and K bands
of Catelan et al. (2004)  are practically identical. 
They also  agree very well with the results obtained 
from the empirical calibration of Sandage (1993) for the V
band, and the V-band data of Kaluzny et al. 1995. This hints at a very small
internal reddening produced inside  the Sculptor galaxy.
The distance modululi obtained from the independent theoretical calibration
of Catelan et al. (2004) and Bono et al.(2003) and empirical calibration
of Sollima et al. (2008) in the K band, are different by some 0.05 mag.
Such a difference is not significant taking into account all
uncertainties involved in those calibrations.

There are several sources of systematic error in our procedure of 
distance determinations. Taking into account the uncertainties of 
the coefficients of the adopted  calibrations, on the one hand; assuming an error of 
the adopted mean metallicity of our RR Lyrae sample
of 0.26 dex as given by Clementini et al. (2005); assuming an accuracy of the 
photometric zero points of 0.02 mag; and finally, taking into account an
uncertainty of 0.01 mag associated to the absorption correction,
 we estimate the systematic errors on each of the distance determinations as listed in Table
4. They are significantly larger than the corresponding statistical 
errors and are dominated by the errors associated with the uncertainty 
of the assumed (constant) metallicity of the individual RR Lyrae stars in our sample.

We adopt
from our study 19.67 $\pm$ 0.02 (statistical) $\pm$ 0.12 (systematic) mag
(e.g. the mean value) as the
true distance modulus of the Sculptor dSph galaxy.
Unfortunately, there is very little information about the  distance to 
Sculptor galaxy in the literature. Our adopted distance agrees very well
with the results obtained by Rizzi (2002) from the optical photometry 
of the TRGB (19.64 $\pm$ 0.08 mag) and HB (19.66 $\pm$ 0.15 mag).

Our results also indicate that the systematic errors of the different 
empirical and theoretical calibrations in the literature
were probably estimated quite conservatively 
and may in fact be  smaller. We will be able to check on this more closely 
once we can compare the present results from IR photometry of RR Lyrae variables
with distance results from other techniques.

29 RR Lyrae stars from our sample are in common with the list of RR 
Lyrae stars with metallicity determinations reported by Clementini et al.
(2005). This provides an opportunity to calculate distances to
individual stars based on equations (1-4) and then  compare the 
results with the values obtained from fitting of the PLK
relations and assuming a mean metallicity of -1.83 dex. The following 
values calculated as the mean from the 29 individual true distance moduli
were obtained: 19.60 $\pm$ 0.05 (Sollima et al., K band), 19.69 $\pm$ 0.05
(Bono et al, K band), 19.66 $\pm$ 0.04 (Catelan et al., K band), and 
19.68 $\pm$ 0.06 (Catelan et al., J band). As can be seen these moduli
are fully consistent with the values listed in the Table 4.

\section{Summary and Conclusions}
We have determined the distance to the Local Group Sculptor dwarf galaxy from
single-phase near-infrared observations in J and K of 78 RR Lyrae stars. Different
calibrations of the near-infrared PLZ relation for RR Lyrae stars yield closely
consistent distance results which also agree with the distance modulus derived from
optical V-band data.

While our data are very well suited for a precise distance determination
to this galaxy, it will be necessary to secure additional data to study in detail
the slope of the PLZ relation in J and K, and to perform a high-precision
empirical determination of the metallicity dependence of the relation.

\acknowledgments
WG and  GP  gratefully acknowledge financial support for this
work from the Chilean Center for Astrophysics FONDAP 15010003. 
Support from the Polish grant N203 002 31/046 is also acknowledged.
It is a special pleasure to thank the support astronomers at ESO-La Silla
for their expert help in the observations, and the ESO OPC for the
generous amounts of observing time at the NTT allocated to our
programme. Finally we would like to thank an anymous referee for 
his/her interesting comments which helped us to improve this 
paper.

\begin{figure}[p] 
\vspace*{18cm}
\includegraphics{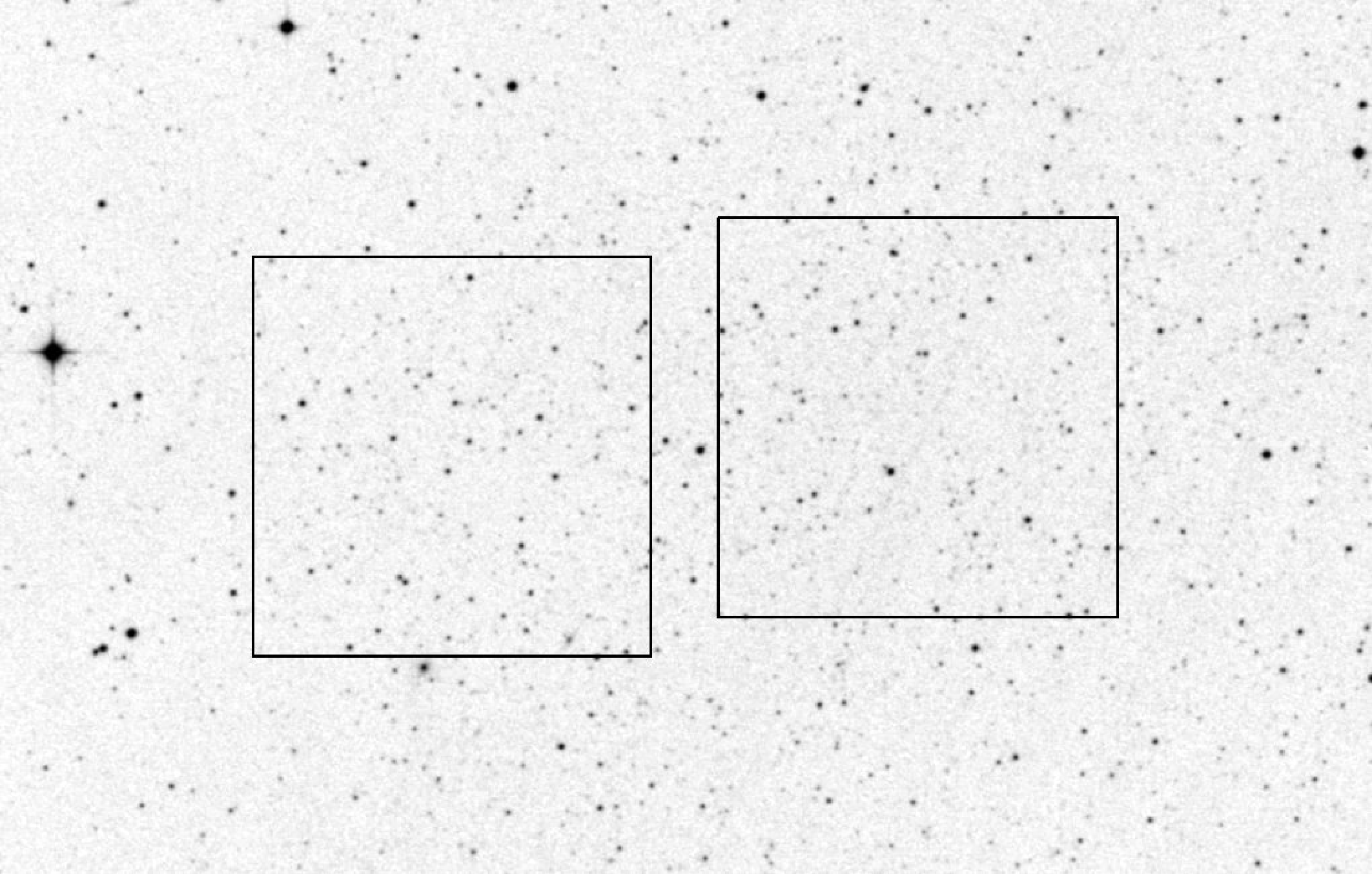} 
\caption{The location of our two observed 5 x 5 arcmin  NTT/SOFI fields 
in Sculptor on the DSS blue plate. North is up and east to the left. 
}
\end{figure}

\begin{figure}[htb]
\vspace*{20cm}
\includegraphics{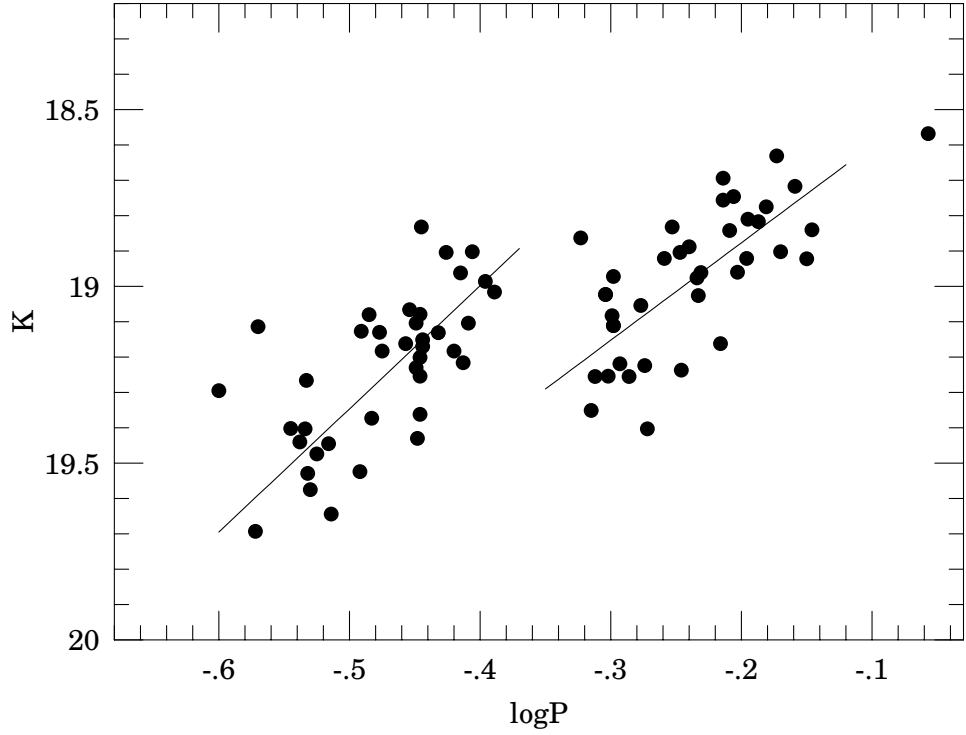}
\includegraphics{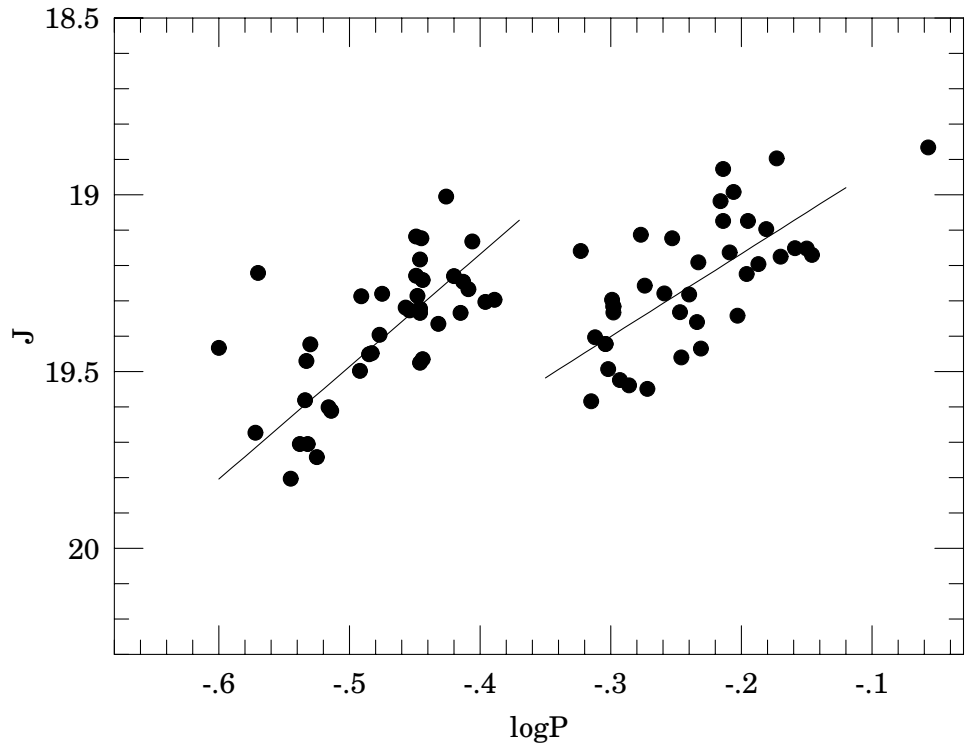}
\caption{The near-infrared K and J band period-luminosity relations 
defined by the 78 RR Lyrae stars observed in Sculptor. Period is in days.
Two sequences formed by
the fundamental and first overtone pulsators are clearly seen in each
panel. The free-fit lines for RRab and RRc stars are indicated. }
\end{figure}

\begin{figure}[htb]
\vspace*{20cm}
\includegraphics{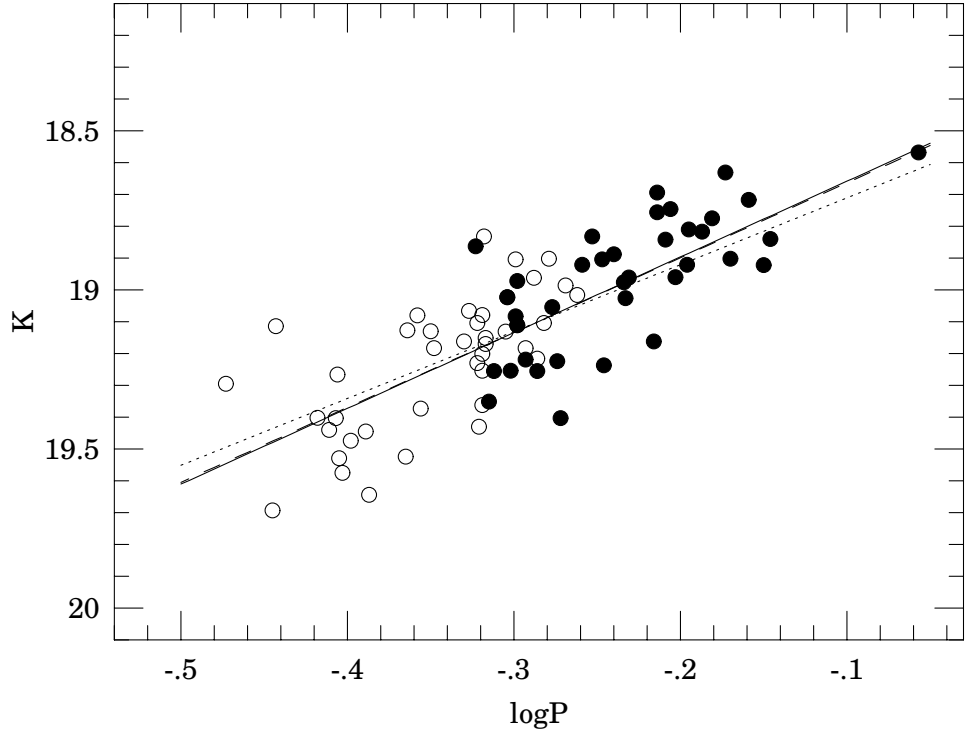}
\includegraphics{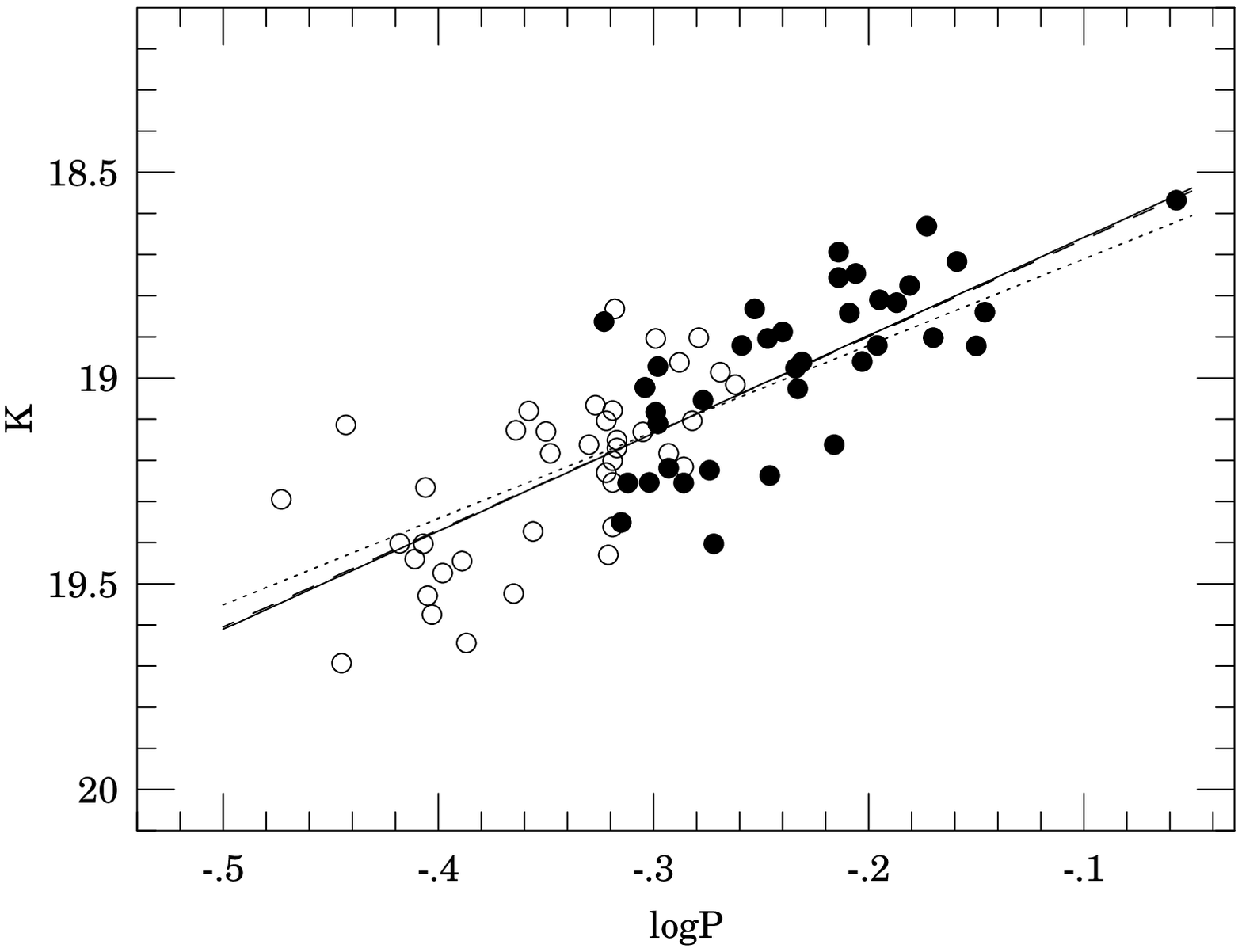}
\caption{
The near-infrared PL relations in K and J defined by our RR Lyrae sample in Sculptor, plotted
along with the best-fitting lines. The filled and open circles 
stand for RRab and RRc stars, respectively. The periods of the RRc stars
have been adjusted by adding 0.127 in logP.
The slopes of the fits were adopted from the recent theoretical and
empirical calibrations (equations 1-4; see text), and the zero points determined
from our data. The solid, dotted and dashed lines correspond to the
calibration of Sollima et al., Bono et al. and Catelan et al., respectively.}
\end{figure}

\clearpage
\begin{deluxetable}{c c c c c c c}
\tablewidth{0pc}
\tablecaption{Journal of the Individual J and K band Observations of 
RR Lyrae stars in Sculptor}
\tablehead{ \colhead{ID} & \colhead{J HJD} & \colhead{J}  & \colhead{$\sigma$} 
& \colhead{K HJD} & \colhead{K} & \colhead{$\sigma$}\\
\colhead{} & \colhead{2400000+} & \colhead{mag} & \colhead{mag} & \colhead{2400000+} &
\colhead{mag} & \colhead{mag}
 }
\startdata
377 & 53371.08459 &   19.42 &    0.06 & 53371.04170 &   19.57 &    0.08 \\
1823 & 53371.08459 &   19.74 &    0.08 & 53371.04170 &   19.47 &    0.08 \\
1824 & 53371.08459 &   19.16 &    0.06 & 53371.04170 &   18.86 &    0.05 \\
1830 & 53371.08459 &   19.54 &    0.07 & 53371.04170 &   19.25 &    0.07 \\
1838 & 53371.08459 &   19.15 &    0.06 & 53371.04170 &   18.92 &    0.05 \\
1873 & 53371.08459 &   19.58 &    0.08 & 53371.04170 &   19.40 &    0.07 \\
1874 & 53371.08459 &   19.22 &    0.06 & 53371.04170 &   19.11 &    0.06 \\
1875 & 53371.08459 &   19.40 &    0.07 & 53371.04170 &   19.13 &    0.07 \\
1877 & 53371.08459 &   19.46 &    0.07 & 53371.04170 &   19.24 &    0.06 \\
1943 & 53371.08459 &   19.28 &    0.08 & 53371.04170 &   18.92 &    0.07 \\
1997 & 53371.08459 &   19.34 &    0.06 & 53371.04170 &   18.96 &    0.06 \\
2004 & 53371.08459 &   19.43 &    0.07 & 53371.04170 &   18.96 &    0.05 \\
2012 & 53371.08459 &   19.17 &    0.06 & 53371.04170 &   18.84 &    0.05 \\
2021 & 53371.08459 &   18.99 &    0.05 & 53371.04170 &   18.75 &    0.05 \\
2048 & 53371.08459 &   19.32 &    0.06 & 53371.04170 &   19.25 &    0.06 \\
2058 & 53371.08459 &   19.32 &    0.06 & 53371.04170 &   19.11 &    0.06 \\
2059 & 53371.08459 &   19.42 &    0.06 & 53371.04170 &   19.02 &    0.06 \\
2410 & 53371.08459 &   19.26 &    0.06 & 53371.04170 &   19.22 &    0.07 \\
2421 & 53371.08459 &   19.30 &    0.06 & 53371.04170 &   19.02 &    0.05 \\
2422 & 53371.08459 &   19.58 &    0.08 & 53371.04170 &   19.35 &    0.07 \\
2423 & 53371.08459 &   19.12 &    0.05 & 53371.04170 &   18.83 &    0.06 \\
2424 & 53371.08459 &   19.32 &    0.06 & 53371.04170 &   19.16 &    0.07 \\
2425 & 53371.08459 &   18.87 &    0.05 & 53371.04170 &   18.57 &    0.04 \\
2450 & 53371.08459 &   19.16 &    0.06 & 53371.04170 &   18.84 &    0.05 \\
2455 & 53371.08459 &   19.22 &    0.06 & 53371.04170 &   18.92 &    0.05 \\
2458 & 53371.08459 &   19.33 &    0.07 & 53371.04170 &   19.36 &    0.09 \\
2467 & 53371.08459 &   19.48 &    0.08 & 53371.04170 &   19.20 &    0.07 \\
2470 & 53371.08459 &   19.15 &    0.06 & 53371.04170 &   18.72 &    0.04 \\
2471 & 53371.08459 &   19.20 &    0.06 & 53371.04170 &   18.82 &    0.05 \\
2482 & 53371.08459 &   19.67 &    0.07 & 53371.04170 &   19.69 &    0.10 \\
2502 & 53371.08459 &   19.45 &    0.06 & 53371.04170 &   19.08 &    0.06 \\
\enddata
\end{deluxetable}

\setcounter{table}{0}
\begin{deluxetable}{c c c c c c c}
\tablewidth{0pc}
\tablecaption{Journal of the Individual J and K band Observations of
RR Lyrae stars in Sculptor - Continuation}
\tablehead{ \colhead{ID} & \colhead{J HJD} & \colhead{J}  &
\colhead{$\sigma$}
& \colhead{K HJD} & \colhead{K} & \colhead{$\sigma$}\\
\colhead{} & \colhead{2400000+} & \colhead{mag} & \colhead{mag} &
\colhead{2400000+} &
\colhead{mag} & \colhead{mag}
 }
\startdata
2552 & 53371.08459 &   19.55 &    0.09 & 53371.04170 &   19.40 &    0.09 \\
2555 & 53371.08459 &   19.30 &    0.07 & 53371.04170 &   19.08 &    0.06 \\
2558 & 53371.08459 &   19.32 &    0.06 & 53371.04170 &   19.11 &    0.06 \\
2559 & 53371.08459 &   19.42 &    0.06 & 53371.04170 &   19.02 &    0.06 \\
2562 & 53371.08459 &   19.25 &    0.06 & 53371.04170 &   19.22 &    0.06 \\
2566 & 53371.08459 &   19.36 &    0.06 & 53371.04170 &   18.98 &    0.05 \\
2575 & 53371.08459 &   18.93 &    0.04 & 53371.04170 &   18.69 &    0.05 \\
2606 & 53371.08459 &   19.19 &    0.07 & 53371.04170 &   19.03 &    0.06 \\
2627 & 53371.08459 &   19.28 &    0.06 & 53371.04170 &   18.89 &    0.05 \\
2639 & 53371.08459 &   19.70 &    0.09 & 53371.04170 &   19.44 &    0.08 \\
3916 & 53371.08459 &   19.60 &    0.08 & 53371.04170 &   19.45 &    0.09 \\
3931 & 53371.08459 &   19.46 &    0.07 & 53371.04170 &   19.17 &    0.07 \\
4233 & 53371.08459 &   19.12 &    0.05 & 53371.04170 &   18.83 &    0.06 \\
4353 & 53371.08459 &   19.24 &    0.07 & 53371.04170 &   19.15 &    0.07 \\
4385 & 53371.08459 &   19.40 &    0.07 & 53371.04170 &   19.25 &    0.08 \\
3761 & 53371.13750 &   19.33 &    0.07 & 53371.09050 &   18.96 &    0.06 \\
3763 & 53371.13750 &   19.50 &    0.09 & 53371.09050 &   19.52 &    0.09 \\
3777 & 53371.13750 &   19.07 &    0.06 & 53371.09050 &   18.81 &    0.05 \\
3801 & 53371.13750 &   19.36 &    0.07 & 53371.09050 &   19.13 &    0.07 \\
3832 & 53371.13750 &   19.29 &    0.09 & 53371.09050 &   19.43 &    0.11 \\
3834 & 53371.13750 &   19.00 &    0.06 & 53371.09050 &   18.90 &    0.06 \\
3862 & 53371.13750 &   19.70 &    0.10 & 53371.09050 &   19.53 &    0.09 \\
3888 & 53371.13750 &   19.02 &    0.05 & 53371.09050 &   19.16 &    0.07 \\
3938 & 53371.13750 &   19.23 &    0.06 & 53371.09050 &   19.18 &    0.07 \\
4235 & 53371.13750 &   19.23 &    0.06 & 53371.09050 &   19.10 &    0.06 \\
4263 & 53371.13750 &   19.80 &    0.11 & 53371.09050 &   19.40 &    0.07 \\
4277 & 53371.13750 &   19.61 &    0.22 & 53371.09050 &   19.64 &    0.20 \\
5330 & 53371.13750 &   19.13 &    0.06 & 53371.09050 &   18.90 &    0.06 \\
5354 & 53371.13750 &   19.18 &    0.07 & 53371.09050 &   18.90 &    0.06 \\
5359 & 53371.13750 &   18.90 &    0.05 & 53371.09050 &   18.63 &    0.05 \\
5375 & 53371.13750 &   19.30 &    0.08 & 53371.09050 &   18.99 &    0.06 \\
5376 & 53371.13750 &   19.27 &    0.07 & 53371.09050 &   19.10 &    0.06 \\
5384 & 53371.13750 &   19.28 &    0.07 & 53371.09050 &   19.18 &    0.08 \\
5390 & 53371.13750 &   19.10 &    0.06 & 53371.09050 &   18.77 &    0.05 \\
\enddata
\end{deluxetable}
                                                                  
\setcounter{table}{0}
\begin{deluxetable}{c c c c c c c}
\tablewidth{0pc}
\tablecaption{Journal of the Individual J and K band Observations of
RR Lyrae stars in Sculptor - Concluded}
\tablehead{ \colhead{ID} & \colhead{J HJD} & \colhead{J}  &
\colhead{$\sigma$}
& \colhead{K HJD} & \colhead{K} & \colhead{$\sigma$}\\
\colhead{} & \colhead{2400000+} & \colhead{mag} & \colhead{mag} &
\colhead{2400000+} &
\colhead{mag} & \colhead{mag}
 }
\startdata

5393 & 53371.13750 &   19.29 &    0.07 & 53371.09050 &   19.13 &    0.07 \\
5400 & 53371.13750 &   19.45 &    0.09 & 53371.09050 &   19.37 &    0.08 \\
5401 & 53371.13750 &   19.33 &    0.08 & 53371.09050 &   18.97 &    0.06 \\
5492 & 53371.13750 &   19.11 &    0.07 & 53371.09050 &   19.05 &    0.07 \\
5710 & 53371.13750 &   19.12 &    0.07 & 53371.09050 &   19.23 &    0.08 \\
5714 & 53371.13750 &   19.47 &    0.08 & 53371.09050 &   19.27 &    0.08 \\
5721 & 53371.13750 &   19.18 &    0.07 & 53371.09050 &   19.08 &    0.07 \\
5723 & 53371.13750 &   19.33 &    0.07 & 53371.09050 &   18.90 &    0.06 \\
5724 & 53371.13750 &   19.49 &    0.09 & 53371.09050 &   19.25 &    0.07 \\
5728 & 53371.13750 &   19.43 &    0.08 & 53371.09050 &   19.30 &    0.08 \\
5730 & 53371.13750 &   19.33 &    0.08 & 53371.09050 &   19.07 &    0.06 \\
5773 & 53371.13750 &   19.52 &    0.09 & 53371.09050 &   19.22 &    0.08 \\
5778 & 53371.13750 &   19.07 &    0.09 & 53371.09050 &   18.76 &    0.08 \\
\enddata
\end{deluxetable}

\begin{deluxetable}{ccccc}
\tablewidth{0pc}
\tablecaption{The Slopes of the K and J band PL relations calculated for 
Sculptor RRa, RRc type stars and the combined sample.}
\tablehead{ \colhead{sample} & \colhead{K} & \colhead{$\sigma$} & \colhead{J} & \colhead{$\sigma$}}
\startdata
RRc  & -2.60 & 0.48 & -2.32 & 0.43 \nl 
RRab & -2.42 & 0.39 & -2.11 & 0.37 \nl
all  & -2.44 & 0.22 & -1.72 & 0.21 \nl
\enddata
\end{deluxetable}

\begin{deluxetable}{ccccc}
\tablewidth{0pc}
\tablecaption{Observed Near-Infrared Distance Moduli for Sculptor
obtained from different theoretical and empirical calibrations.}
\tablehead{ \colhead{} & $(M-m)_{\rm K}$ & $(M-m)_{\rm K}$ & $(M-m)_{\rm K}$  &
$(M-m)_{\rm J}$ \\
Calibration & Sollima et al. & Bono el al. & Catelan et al. & Catelan et al.}
\startdata
RRab                         &   19.641 $\pm$ 0.023 &  19.712 $\pm$ 0.023 & 19.669 $\pm$ 0.023 &  19.684 $\pm$ 0.022\nl
RRab + RRc                   &   19.643 $\pm$ 0.017 &  19.727 $\pm$ 0.017 & 19.686 $\pm$ 0.017 &  19.686 $\pm$ 0.017\nl
\enddata
\end{deluxetable}

\begin{deluxetable}{cccccc}
\tablewidth{0pc}
\tablecaption{Reddening-corrected  V, J and K Band Distance Moduli for Sculptor
obtained from RR Lyrae stars and using different theoretical and empirical calibrations. 
The reddening law of Schlegel et al (1998) was assumed, and the 
foreground reddening E(B-V) = 0.018 toward the Sculptor galaxy was calculated
using reddening maps from this same paper.}
\tablehead{ \colhead{Filter} & K & K & K & J & V }
\startdata
data        & this paper & this paper & this paper & this paper &
Kaluzny et al. (1995) \nl
Calibration & Sollima et al. & Bono el al. & Catelan et al. & Catelan et
al. & Sandage 1993 \nl
$(m-M)_{0}$ &  19.64 &  19.72  & 19.68 &  19.66 & 19.67\nl
statistical error &  0.02 & 0.02 & 0.02 & 0.02 & 0.08\nl
systematic error  & 0.15 & 0.13 & 0.12 & 0.12 & 0.18 \nl
\enddata
\end{deluxetable}

\end{document}